\documentclass[reprint,nofootinbib,longbibliography,...]{revtex4-1}

\pdfoutput=1

\usepackage{amsmath}
\usepackage{graphicx}
\usepackage{mathtools}
\usepackage{amssymb}
\usepackage[utf8]{inputenc}
\usepackage[english]{babel}
\usepackage{amsthm}
\usepackage{bbm}
\usepackage{subcaption} 
\usepackage[titletoc,title]{appendix}
\usepackage{url}
\usepackage{lmodern}

\theoremstyle{plain}
\newtheorem{theorem}{Theorem}

\newtheorem{lemma}[theorem]{Lemma}
\newtheorem*{lemma*}{Lemma}

\theoremstyle{definition}
\newtheorem{definition}[theorem]{Definition}

\begin{document}

\title{Consistency and convergence of simulation schemes in information field dynamics}
\author{Martin Dupont}
\affiliation{Max Planck Institute for Astrophysics, Karl-Schwarzschild-Str. 1, 85741 Garching, Germany}
\author{Torsten En{\ss}lin}
\affiliation{Max Planck Institute for Astrophysics, Karl-Schwarzschild-Str. 1, 85741 Garching, Germany}

\begin{abstract}
We explore a new simulation scheme for partial differential equations (PDE's) called Information Field Dynamics (IFD). 
Information field dynamics is a novel probabilistic numerics method  that seeks to preserve the maximum amount of information about the field being simulated. 
It rests on Bayesian field inference and therefore allows the incorporation of prior knowledge on the field.
This makes IFD attractive to address the closure problem of simulations; how to incorporate knowledge about sub-grid dynamics into a scheme on a grid with limited resolution.
Here, we analytically prove that a restricted subset of simulation schemes in IFD are consistent, and thus deliver valid predictions in the limit of high resolutions. 
This has not previously been done for any IFD schemes. 
This restricted subset is roughly analogous to traditional fixed-grid numerical PDE solvers, given the additional restriction of translational symmetry. 
Furthermore, given an arbitrary IFD scheme modelling a PDE, it is a priori not obvious to what order the scheme is accurate in space and time. 
For this subset of models, we also derive an easy rule-of-thumb for determining the order of accuracy of the simulation. 
As with all analytic consistency analysis, an analysis for nontrivial systems is intractable, thus these results are intended as a general indicator of the validity of the approach, and it is hoped that the results will generalize. 

\end{abstract}

\maketitle

\section{Introduction}
\subsection{Probabilistic numerics}
Information Field Dynamics (IFD) is a new framework for constructing numerical simulation schemes for partial differential equations (PDE's) that can be cast into the form
\begin{equation}
	\partial_t\, \phi = F[\phi]\label{eq:PDE}
\end{equation}
with $\phi$ being some field, and $F$ some linear or non-linear endomorphic operator in the Hilbert space of field configurations. 
IFD was first proposed in \cite{IFD}, and further developed in \cite{IFDMath} and \cite{ReimarTowards}. 
%
%
IFD addresses the problem of constructing numerical simulation schemes probabilistically. 
It therefore belongs to the emerging field of \textit{probabilistic numerics} \citep{HenOsbGirRSPA2015}.\footnote{See also \url{http://probabilistic-numerics.org}.}

In classical numerics, a point estimate of the quantity of interest is provided.
In contrast to this, in probabilistic numerics, a probability distribution over possible values of the quantity of interest is constructed and investigated.
This permits the quantification of uncertainties on the results as well as the incorporation of domain knowledge into their estimation.
%
%

Classical numerical simulation schemes therefore follow a representation of the evolving field in time. 
This is done by updating some data in computer memory, which, for example, specifies the field values within the voxels of a discretized field domain. 
The field configuration and its discretized representation are the central elements of classical simulation schemes.
In probabilistic numerics, the central object is not the field itself, but the probability of the field having a specific configuration.
The basic idea of IFD is therefore to follow the evolution of the probability distribution of possible field states. 
The data in computer memory of a probabilistic simulation scheme therefore parametrizes this probability distribution.
This distribution can be used to ask questions, such as: what is the most probable field configuration, what is the mean field, or what is its uncertainty covariance?

Some of the earliest works in the field of probabilistic numerics are \cite{Diaconis, oHagan, Larkin}, who treated the problem of function interpolation as a statistical inference problem. 
For the study of differential equations, \cite{Skilling1}, was the first to propose treating ordinary differential equations (ODE's) as a Bayesian inference problem.
The papers \cite{Lecot, Stengle} were early examples of using randomized Monte-Carlo methods for the solution of ODE's.
More recently, \cite{Briol} used Bayesian uncertainty to quantify and reduce errors in numerical integration, and \cite{Schober} develops a Gaussian process solver which generates a probability distribution over a set of solutions to an ODE, which are centred around a Runge-Kutta solution.
The paper \cite{Conrad} is particularly relevant for our work, as it analyzes the convergence properties and errors of probabalistic solvers for PDE's. 

Our proposed scheme is somewhat different to previous works, as our algorithms are deterministic, although they nonetheless incorporate a notion of prior belief, measurement, and uncertainty.
One paper which proposes a scheme similar to IFD is \cite{Archambeau}, which solves stochastic differential equations by incorporating prior beliefs, and a notion of observations, and like our work, derives its equations of motion using the criteria of minimal Kullbach-Leibler divergence. 

\subsection{Minimal information loss}
The dynamical equation that governs the evolution of the field $\phi$ determines how the probability distribution of field configurations should evolve. 
IFD attempts to follow the evolution of this probability distribution as a whole. 
Since the representation of this distribution is parametrized and therefore can not perfectly represent any resulting distribution, an approximation scheme is required to map the evolved distribution back into the space of distributions that can be represented by the computer data. 
We take it as a principle that this approximation should conserve as much of the information of the full distribution as possible.

An appropriate measure of the amount of information lost in this approximation is the Kullbach-Leibler (KL) divergence of the approximated distribution $\mathcal{Q}(\phi)$ with respect to the more accurate distribution (or measure) $\mathcal{P}(\phi)$ \citep{Reimar},
\begin{equation}
	\mathrm{KL}(\mathcal{P}||\mathcal{Q})=\int \mathcal{D}\phi\, \mathcal{P}(\phi)\,\ln \left(\frac{\mathcal{P}(\phi)}{\mathcal{Q}(\phi)}\right).
\end{equation}
Here, $\mathcal{P}(\phi)$ stands for the time-evolved probability of field configurations, and $\mathcal{Q}(\phi)$ for the parametrization chosen to represent this approximatively.
The more mathematically inclined reader should read the path integral $\int \mathcal{D}\phi\, \mathcal{P}(\phi)$ as $\int d\mathcal{P}(\phi)$, an average over $\phi$ with the measure $\mathcal{P}(\phi)$.

The KL is now minimized with respect to the parameters of $\mathcal{P}$. 
This then provides an update rule for these parameters in computer memory, which then represent the desired simulation scheme. 

This scheme evolves a probability distribution via its chosen representation in computer memory.
The action principle for this is given by the requirement of minimal information loss, ensuring that the simulation is as accurate as possible.
The precise details of the derivation will be spelled out in the following sections. 

\subsection{Information field theory}
In principle, any suitable parametrization of the field probability distribution could be chosen and the field dynamic mapped to it via the principle of minimal information loss.
In information theory for fields, or \textit{information field theory} (IFT) for short \cite{2018arXiv180403350E,IFT,Lemm}, a canonical representation of field probabilities as a function of some data already exists.

IFT was designed to address the problem of field inference from measurement data.
It turns the data $d$ into a posterior probability $\mathcal{P}(\phi|d)$ on the field $\phi$, which was measured. 
Thus, IFT provides us with a convenient parametrization of field probabilities, which depend on data. 
Now, IFD uses this parametrization and therefore just regards the data in computer memory as the result of a virtual measurement process (more on this later).
The process can be chosen arbitrarily, however it should ideally provide an analytically tractable posterior distribution $\mathcal{P}(\phi|d)$.

For this reason, a simple linear measurement equation of the form
\begin{equation}
	d=R\,\phi +n \label{measurement}
\end{equation}
will be chosen for this paper. Here, $R$ represents the measurement response, an operator that maps the continuous field into a finite-dimensional data vector, and $n$ is some field-independent Gaussian random noise vector with known covariance $N=\langle n\,n^\dagger \rangle_{(n)}$. 

As the data is finite, but the field has infinitely many degrees of freedom, field inference is usually an ill-posed problem that requires regularization, i.e. the removal or suppression of implausible solutions that are otherwise allowed by the data.
This is provided by the field prior, which for convenience assumes the field to be drawn from a zero-centered Gaussian process with covariance $\Phi=\langle \phi\,\phi^\dagger \rangle_{(\phi)}$,
\begin{equation}
	\mathcal{P}(\phi)=\mathcal{G}(\phi,\Phi)\equiv \frac{1}{|2\pi\Phi|^\frac{1}{2}}\,\exp\left(-\frac{1}{2}\phi^\dagger \Phi^{-1} \phi \right),
\end{equation}
where $\phi^\dagger \psi \equiv \int dx\,\overline{\phi(x)}\,\psi(x)$
denotes the canonical scalar product of the Hilbert space\footnote{In this paper we will refer to this Hilbert space as \textit{field space} and the vector space in which the data resides will be called \textit{data space}.}.
Bayes' theorem then provides the field posterior,
\begin{equation}
	\mathcal{P}(\phi|d) = \frac{\mathcal{P}(d|\phi) \, \mathcal{P}(\phi)}{\mathcal{P}(d)},
\end{equation}
where $\mathcal{P}(d|\phi)=\mathcal{G}(d-R\phi,N)$ is the likelihood; the probability of the obtained data $d$ given a field configuration $\phi$.
This field posterior turns out to be a Gaussian
\begin{equation}
	\mathcal{P}(\phi|d)=\mathcal{G}(\phi-m,D)
\end{equation}
under the simplifying assumptions made here \cite{IFT}. 
The posterior mean
\begin{equation}
	m = W\,d \equiv D\,R^\dagger N^{-1}\,d
\end{equation}
is a linear function of the data, as $m=m(d)$. 
In contrast to this, the posterior uncertainty dispersion
\begin{equation}
	D = \left(\Phi^{-1} + R\,N^{-1} R\right)^{-1} 
\end{equation}
is independent of the data for this linear and Gaussian field estimation problem .
The operator $W$ turning the data into the posterior mean field is called Wiener filter in signal processing and $D$ is also called the Wiener covariance.
In this paper, we will often express this operator in its so called data space version\footnote{This is so named, since here the operator inversion happens in data space.}
\begin{equation}
	W = \Phi\,R^\dagger \left(R\,\Phi\,R^\dagger + N \right)^{-1},
\end{equation}
as this form allows us to take the no-noise limit $N\rightarrow 0$, in which $	W \rightarrow \Phi\,R^\dagger \left(R\,\Phi\,R^\dagger \right)^{-1}$.
IFT also extends this linear signal inference to non-linear and non-Gaussian problems. 
However, for IFD as developed so far, this is not needed and the Wiener filter theory presented here is sufficient.

\subsection{Information field dynamics}
IFD regards the data as being in some sense a measurement of the field being simulated.
This idea can be taken literally or not.
For example, using the formalism one could take the initial data to be the result of a literal measurement, and IFD would prescribe a way of simulating how future measurements of the time-evolved field would appear.

However, one can also regard the response $R$ as simply being a mathematical object which creates a finite-dimensional representation of the continuous field being simulated, and the noise just represents some degree of uncertainty in this description.
If, for example, one defines a response that takes samples of the field at particular points, and sets the noise to zero, then the data becomes identical to the gridpoints of a typical finite-difference scheme.
Thus, the idea of representing data as a virtual measurement of a field being simulated should not be too unusual. 
The point is, that in order to run a simulation, one needs to express a field using a finite amount of information, and the formalism of IFT provides us with a convenient way of doing so.

In an IFD simulation, there are many parameters that may be updated in time to best capture the time evolution of the posterior distribution.
These include the data $d$, but potentially also the properties of the measurement equations, the response $R$ and the noise covariance $N$, as well as the field prior covariance $\Phi$.

Changing the data $d$ while keeping all other parameters constant corresponds most closely to a typical finite-difference simulation scheme; the data points represent samples of a field, and these change in time as the simulation progresses. 
A response $R$ which changes in time would be analogous to changing the coordinate system during a simulation in order to best capture the behaviour of the field under consideration.
In this paper, we will only consider schemes where only the data is updated. 

We now present an abridged derivation of the IFD simulation scheme. We restrict ourself to linear dynamics:
\begin{equation}
\partial_t\, \phi = L\,\phi \label{eq:lPDE}
\end{equation}
with $L$ some linear, time-independent, endomorphic operator.
This equation has a formal solution given by $\phi(t)=U(t)(\phi_0)$, for 
$U(t)=\exp (tL)$. 

The scheme is then as follows: It is assumed that there is some data $d_i$ taken at some point in time, $t_i$, which is interpreted as being some coarse-grained representation of the true field $\phi(t_i,x)$, which is obtained by some linear measurement as in eqn.~(\ref{measurement}). 
To run the simulation, i.e. obtain the data at $t_{i+1}$ from that at $t_i$, IFT is used to reconstruct the posterior probability distribution of the field, $\mathcal{P}(\phi(t_i,x)|d_i)$, given the initial data, $d_i$. 

This posterior distribution is then evolved from $t_i$ to $t_{i+1}$ using the equations of motion for the field.
This would formally be done using $U(\Delta t)$.
However, to achieve a practical simulation scheme, the time evolution $U$ must be truncated to some finite order\footnote{If we could write down a closed-form expression for $U$, then we wouldn't need to run the simulation.}, which we denote by $\bar{U}=\sum_{k=0}^{\alpha} (\Delta t L)^k/k!$ for some order $\alpha$, which corresponds to a choice of the desired time-order accuracy of the simulation.

To obtain the data at the next timestep, a second measurement, $R_{i+1}$,  is postulated, which is used to construct a second posterior distribution. 
The new data is then chosen as to minimize the information loss between the evolved and unevolved posterior distributions, using the Kullback-Leibler divergence (KL divergence). With linear dynamics, measurements and a Gaussian prior\footnote{We refer to this as the \textit{linear case} of IFD.}, the resulting finite-difference equation becomes particularly transparent \cite{ReimarTowards}:
\begin{equation}\label{dataupdate}
\boxed{d_{i+1}=(R_{i+1}W_{i+1})^{-1}R_{i+1}\bar{U}W_i d_i}
\end{equation}
The subscripts denote time indices, as the response and prior covariance are allowed to vary between timesteps.  
The above equation is in the form of a matrix equation, and the matrix will be referred to as the difference operator and will be denoted by $T_i=(R_{i+1}W_{i+1})^{-1}R_{i+1}UW_i$.

\begin{figure}
\includegraphics[width=\linewidth]{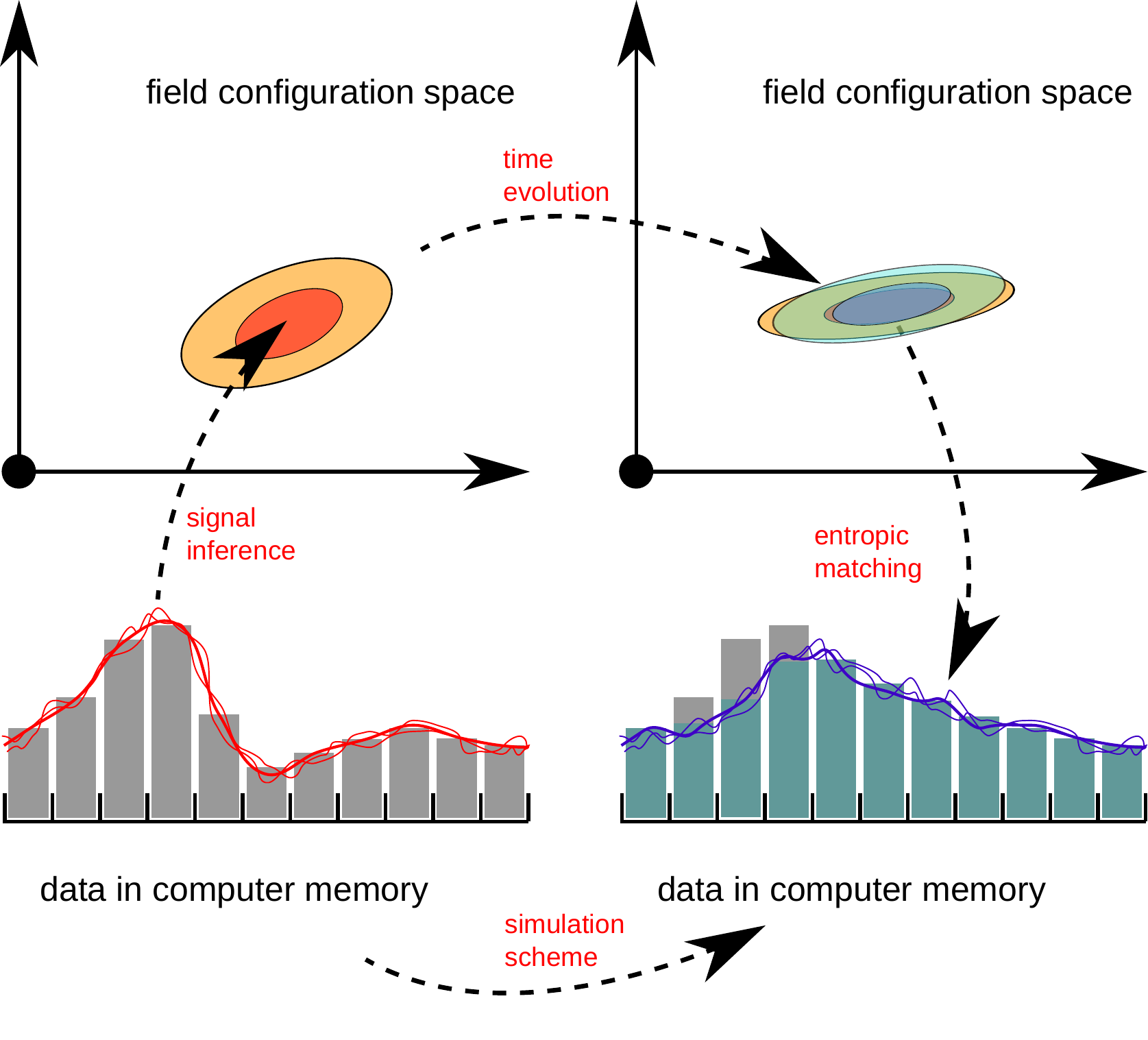}
\caption{Schematic representation of the IFD concept.
	Data in computer memory (gray bars) imply via Bayesian signal inference a posterior probability distribution in field space (orange density contours on the left). 
	For each field configuration (each point of this space) the time evolution assigns it a new location. 
	The time-evolved probability distribution (orange contours on the right) needs to be re-expressed in terms of new data.
	This is done via entropically matching this (or minimizing the KL) to a new posterior distribution (blue contours) expressed by new, time evolved data (blue bars).
	The set of implied operations on the data represents the IFD simulation scheme. 
	It incorporates the field dynamics, prior knowledge on the field (e.g. on sub-grid scales), and tries to conserve as much information on the field as possible.	
	}
\label{fig:IFD7}
\end{figure}

A new development, which we present in this paper, is that this difference equation may be simplified further in the linear case, because repeated applications of the difference operator during the simulation will result in certain useful cancellations. The result of which, is that we may always assume that we are operating in the no-noise limit.  

\begin{lemma}
The finite-difference equations for linear IFD are independent of the noise up to a simple equivalence.
\begin{proof}
For a simulation scheme with timesteps ${t_i}$ for $i \in \{1,...n\}$, responses $\{R_i\}$, priors $\{ \Phi_i\}$, noises $\{ N_i\} $, Wiener filters $\{ W_i=\Phi_i R_i^\dagger (R_i \Phi_i R_i^\dagger +N_i)^{-1}\}$, and linear time evolution operators $\bar{U}_i= \mathbbm{1}+\Delta t L_i+...$, the finite-difference equation is given by:
\begin{multline}
d_{i+1}=(R_{i+1}W_{i+1})^{-1}R_{i+1}\bar{U}_i W_i d_i \\
=\big{[}R_{i+1}\Phi_{i+1} R_{i+1}^\dagger (R_{i+1}\Phi_{i+1} R_{i+1}^\dagger +N_{i+1})^{-1}\big{]}^{-1} \\
\cdot R_{i+1}\bar{U}_i \Phi_{i} R_{i}^\dagger (R_{i}\Phi_{i} R_{i}^\dagger +N_i)^{-1}d_i.
\end{multline}

The second line is obtained by inserting the data-space definition of the Wiener filter. 
We rename the terms: $(R_i \Phi_i R_i^\dagger +N_i)=C_i$ , $(R_i \Phi_i R_i^\dagger)=B_i$ and $R_{i+1} \bar{U}_i \Phi_i R_i^\dagger=A_i$,
yielding:
\begin{equation}
d_{i+1}= (B_{i+1} C_{i+1}^{-1})^{-1} A_i C_{i}^{-1} d_i =  C_{i+1} B_{i+1}^{-1} A_i C_{i}^{-1}d_i.
\end{equation}
The difference equations are then iterated $n$ times. With every matrix multiplication, there is a $C_{i}$ which cancels with a $C_{i}^{-1}$, yielding:

\begin{equation}
d_n= C_n \big ( \prod_{i=0}^{n} B_{i+1}^{-1} A_i \big ) C_0^{-1}d_0.
\end{equation}

The only noise-dependent terms were the $C$ terms and therefore, up to a change of basis at the beginning and end of the simulation, the difference equations are independent of the noise. 
In the infinite-noise limit, $C \to N$, and in the zero noise limit $C \to B$.

\end{proof}
\end{lemma}

Given the equivalence, from here on we will always work in the no-noise limit, and the symbol $N$ will be used to denote number quantities. 
In this limit, the difference operator becomes: 
\begin{equation}\label{updateOperator}
T_i=R_{i+1} \bar{U}_i \Phi_i R_i (R_i \Phi_i R_i^\dagger)^{-1}
\end{equation}

Although these incarnations of IFD schemes, eqs.~(\ref{dataupdate}), and~(\ref{updateOperator}), might intuitively make sense, it still is not guaranteed that they lead to consistent and convergent numerical simulations.

\subsection{Numerical consistency and error}

The major goal of this paper is to show that in a restricted setting, the IFD equations are \textit{consistent}. 
This is a valuable goal, for the Lax Equivalence theorem 
\cite{Lax} states that if a scheme is consistent, then it converges to the true solution if and only if it is \textit{stable}. 
We state the (paraphrased) definition of consistency:

\begin{definition}[Consistency] \label{cons}
For an operator $T(\Delta t,\Delta x)$ which approximates $U(t)$, with $U(t)$ being the analytic time evolution operator corresponding to $L(t)$, 
the approximation is said to be consistent, if for some set of genuine\footnote{See the original publication for a definition of a genuine solution.} solutions $\Omega$ to the differential equation, then for any $\phi \in \Omega$, 
\begin{equation}\label{consEquation}
\lim_{\Delta t,\Delta x \to 0} \bigg{\|} \big{(} T(\Delta t,\Delta x) - U(\Delta t) \big{)} \phi(t,x) \bigg{\|} =0
\end{equation}
uniformly in $t$.
\end{definition}

Note that the above definition involves comparing operators which are defined on different spaces: $T(\Delta t,\Delta x)$ acts on a discrete space, yet $U(\Delta t)$ acts on a continuous space. 
Ref. \cite{Lax} assumes that there is some sufficient level of smoothness such that Taylor series expansions or smooth interpolation etc. 
may be used to approximate the norm. 
We discuss a way of comparing these two operators in IFD later, once a more concrete expression for $T$ has been found. 

The other goal of this paper is to analyze the numerical error of IFD schemes,  and how such error scales as the spatial and temporal resolutions $\Delta x$ and $\Delta t$ become arbitrarily fine. 
IFD is a nominally information-theoretic framework, so it is conceivable that one could try and use some information-theoretic notion of error.
However, this would limit our ability to compare the performance of IFD schemes to standard finite-difference schemes. 
Thus, in this paper, we follow a standard approach and analyze the \textit{local truncation error}, or the \textit{one-step error} \cite[p.593]{Numerics}.
We do this by analyzing the distance in the operator norm of the difference operator and the true analytic time evolution operator.
This distance provides a bound on the error which can accumulate during a single timestep of the simulation. 

\section{Argumentation}
\subsection{Groundwork}

We now begin the work of proving consistency by defining the type of models we will be working on. 
We restrict our focus to PDE's for which $L$ is translation-invariant. 
This case can already be solved analytically by Fourier analysis. 
However, this practice is entirely normal in numerical methods, as many advanced simulation schemes are too complicated to permit an analytic analysis \cite[ch.~7]{Godunov}. 
Such is the case in IFD; as the codes are typically nonlocal, meaning the algebraic equations tend to be dependent on the global geometry of the simulation domain. 
Thus, the best that one can do is prove convergence for the analytically solvable case, and then hope that these conclusions hold in the non-analytically solvable case.  
For pedagogical clarity, the results presented in this section are for one spatial dimension only, although it is argued later that they generalize trivially.

We first restrict ourselves to the case where the response $R$ and prior $\Phi$ do not change in time, i.e. the coordinate system is static and our prior belief about the system will not change during the simulation. 
Because these quantities are now static, there is no need to subscript them to denote the timestep in question. This allows us to free up the subscripts for other purposes.

We now select the field and data spaces. 
The simulated space inside the computer must always be of finite extent. 
For this reason, we choose the field space to be $\mathcal{L}^2([0,l])$. 
We apply periodic boundary conditions to render the analytic equations tractable. 

Now the prior must be selected. 
If the PDE under consideration is translation-invariant, then one should choose a prior belief which is also translation-invariant.
Thus the prior covariance will have a diagonal representation in Fourier space. 
The positivity and self-adjointness conditions on the prior covariance ensure that the eigenvalues in momentum space will be everywhere positive and greater than zero, and symmetric about the origin. 
Priors of this form are generally referred to as smoothness priors. 
Using $k$ to denote momentum, a prior $\Phi_{kk}$ whose values fall to zero as $k \to \infty$ essentially states that rapid oscillations in the signal are deemed unlikely; the field is smooth.
For notational convenience, we will often denote the diagonal entries of the prior, $\Phi_{kk}$ as $\Phi(k)$.

Simple examples of a prior include power laws in momentum, i.e. 
$|k|^{-\beta}$ for some integer $\beta$, often supplemented by a regularizing mass term: ${\Phi(k) = 1/(|k|^\beta + m^\beta)}$. 

We pick the responses by assuming that we have $N$ spatial points which will be labelled with the index $j$. 
The responses are chosen to be constant in time, and the subscripts $R_j$ now denote spatial indices. 
The most natural and naive response is to choose the index $j$ to label a regular grid of positions. 
We define $\Delta x=l/N$. 
We let the response be any response which measures the field by integrating over some function $B(x)$ on $\mathcal{L}^2([0,l])$ localized at the point $x_j$:

\begin{equation}
(R \phi)_j= \int_0^l dx B(x-x_j)\phi(x)
\end{equation}

\noindent where $x_j$ is the $x$-position of the $j$-th gridpoint, i.e. 
$x_j = \Delta x\cdot j$. 
A simple example of such a function could be the box function:

\begin{equation}\label{boxExample}
B(x)=
\begin{cases} 
      1/{\Delta x} & 0\leq x\leq \Delta x \\
      0 & $otherwise$.
   \end{cases}
\end{equation}
The response is then an average of the field around that point. 
If the $x_j$'s are evenly spaced, we refer to any response of this form as a \textit{translation-invariant response}. 
The $B(x)$ functions will be referred to as the \textit{response bins} or just \textit{bins}. 

We now begin to calculate the difference operator, starting with the computation of $(R \Phi R^\dagger)^{-1}$. 
Since both the responses and prior covariance are invariant under translations of multiples of $\Delta x$, we can make a very general statement:

\begin{lemma}\label{translation}
Given a field space of the form $\mathcal{L}^2([0,l])$ with periodic boundary conditions, a translation invariant response $R_j$ whose bin function $B(x)$ has a Fourier series representation, as well as a prior covariance $\Phi$ which is diagonal in momentum space, $(R \Phi R^\dagger)_{jl}$ will be of the form:
\begin{equation}
\sum_{k}^{} \Phi(k) |\widehat{B}(k)|^2 e^{ik(x_j-x_l)},
\end{equation}
where $\widehat{B}(k)$ is the Fourier coefficient of $B(x)$.
\begin{proof}
By the shift property of the Fourier transform, $\widehat{R}_{j,k}=e^{-ikx_j} \widehat{R}_{0,k} = e^{-ikx_j}\widehat{B}(k)$. 
Therefore $(R \Phi R^\dagger)_{jl}$ is
\begin{align}
(R\Phi R^\dagger)_{jl} &= \sum_{k}^{} \sum_{q}^{} e^{ikx_j}\widehat{B}(k) \Phi_{kq} \widehat{B}^{*}(q)e^{-iqx_l} \nonumber \\
&=\sum_{k}^{} \Phi(k) |\widehat{B}(k)|^2 e^{ik(x_j-x_l)}
\end{align}
as desired. 
\end{proof}

\end{lemma}

\noindent It must be stressed that we are not demanding that the simulation is carried out in Fourier space, we are rather stating that the operator will always have such a representation. 
From now on, any simulation scheme which satisfies the criteria of the previous lemma, and in addition has a translation-invariant time evolution operator $U$, will be referred to as a \textit{translation invariant scheme}.

The $R\Phi R^\dagger$ matrix must now be inverted, however the inverse is not equal to the inverse of the Fourier coefficients.
Observe that the spatial gridpoints are both finite \textit{and} discrete, which means 
that terms such as $ \sum_{j}^{} e^{ix_j(k-q)}$ do not form Kroenecker deltas $\delta_{kq}$. 
The sum equals $N$, not only when $k=q$ but also when $\frac{l}{2\pi N}(p-q)$ is an integer. 

The reason for this is that data space is a discrete periodic interval which has a discrete Fourier transform (DFT). 
For a DFT, the momentum values $k$ are the same as those for the continuous interval, albeit with a highest uniquely resolvable frequency known as the \textit{Nyquist frequency}, which is equal to half of the sampling frequency. 
In this case, the Nyquist frequency is $\frac{\pi} {\Delta x}$ and is denoted by $f_{N}$. 
Given that the matrix is indeed translation-invariant in data space, it must have \textit{some} diagonal representation in the discrete Fourier transform, i.e. some scalar function of $k$, for $k$ now less than $f_N$. 
This representation can be found by resumming over multiples of the Nyquist frequency:

\begin{lemma}\label{resum}
Given a regular, discrete grid of points $\{ x_j \}$ for $j \in \{ 1, \ldots, N \}$ on a periodic interval, and a matrix of the form:

\begin{equation}\label{brillouin}
A_{lj}=\sum^{\infty}_{k=-\infty} f(k)e^{ik(x_l-x_j)}
\end{equation}
for $f(k)$ some function of $k$, it has a diagonal representation in the DFT Fourier space, given by:

\begin{equation}
A_{lj}= \sum^{f_N}_{|k|} \underbrace{\bigg ( \sum^{}_{b \in 2f_N \mathbb{Z}}f(k+b) \bigg)}_{\equiv g(k)} e^{ik(x_l-x_j)} =\sum^{f_N}_{|k|} g(k) e^{ik(x_l-x_j)}.
\end{equation}

\begin{proof}
We partition the infinite sum over $k$ in eqn.~(\ref{brillouin}) into smaller sums shifted by multiples of the Nyquist frequency.\footnote{Note that depending on whether the number of data points is even or odd, the domain of $|k|< f_N$ changes. 
For odd $N$ we use the convention that $k \in [-(N-1)/2, (N-1)/2]$ and if it's even we use $ k \in [-N/2, N/2 -1]$.} For 
any $x_i$ and $x_j$ separated by a multiple of $\Delta x$ and $b = 2 \pi n/ \Delta x$, we have $(k+b)(x_i-x_j) = k(x_i-x_j) + 2\pi n$. 
This factor of $2 \pi$ then 
disappears in the complex exponential, yielding the desired result:

\begin{align}
A_{lj}=\sum^{< f_N}_{|k|} \sum^{}_{b \in 2f_N \mathbb{Z}}f(k+b)e^{i(k+b)(x_l-x_j)} \\
\stackrel{\text{Nyquist} }{=} \sum^{f_N}_{|k|} \bigg ( \sum^{}_{b \in 2f_N \mathbb{Z}}f(k+b) \bigg) e^{ik(x_l-x_j)}.
\end{align}

This resummed function is a diagonal function of the DFT frequencies $k < f_N$, and so must be the desired operator. 
\end{proof}
\end{lemma}

\noindent

Due to the physical analogy with Brillouin zones, we refer to the procedure of summing over multiples of the Nyquist frequency as \textit{the sum over Brillouin zones}.

Now that we have obtained a representation of the operator which is diagonal in the DFT space, inverting follows easily by taking the inverse of the DFT Fourier coefficients: 

\begin{equation}
(R \Phi R^\dagger)_{lj}^{-1}= \frac{1}{N} \sum^{f_N}_{|k|} \frac{e^{ik(x_l-x_j)}}{\sum^{}_{b \in 2f_N \mathbb{Z}}\Phi(k+b) |\widehat{B}(k+b)|^2} .
\end{equation}

The factor of $N$ comes from the different normalizations of the DFT and the regular Fourier transform. 
Fourier modes in the DFT are normalized as $\frac{1}{\sqrt{N} }e^{-ikx_j}$, as opposed to $\frac{1}{\sqrt{l}}e^{-ikx}$ for the continuous Fourier series. 

It is now time to compute the second part of the difference operator,  $R\bar{U}\Phi R^\dagger $. 
Given that $L$ is assumed to be translationally-invariant, $\bar{U}$ will have a diagonal representation in Fourier space. Thus, the previous lemma~(\ref{translation}) applies, and the operator will also be diagonal in the DFT space, with a sum over Brillouin zones. 
With this information, we may now write down the general form of the difference operator $T= R\bar{U}\Phi R^\dagger (R\Phi R^\dagger)^{-1}$ for translation-invariant systems:

\begin{equation} \label{main}
T_{lj}= \sum^{f_N}_{|k|} 
\frac{ \sum^{}_{b \in 2f_n \mathbb{Z}} \bar{U}(k+b) \Phi(k+b) |\widehat{B}(k+b)|^2}
{\sum^{}_{\hat b \in 2f_n \mathbb{Z}} \Phi(k+\hat b) |\widehat{B}(k+\hat b)|^2}e^{ik(x_l-x_j)}.
\end{equation}

The factor of $1/N$ is cancelled by a factor of $N$ coming from the sum over spatial indices. 

\subsection{Consistency}
The main objectives of this paper are concerned with the behaviour of the difference operator in the limit of high spatial and temporal resolutions.
Given that the spatial points are not just localized samples of the field, but rather the outputs of a response operator $R$, it is not immediately clear how to take a limit of spatial resolution.
The number of points $x_i$ could be increased, although as the number of points approaches infinity, the data would not begin to look like the true field, as the shape of the bin functions has not changed.
Thus, we propose that the correct way to increase the spatial resolution is to increase the number of bins, and decreasing the width of the bin functions such that they approach something resembling delta functions in the limit. 
Rigorously formulated, the process is as follows: given some initial resolution $\Delta x_0$ for which the points $x_i$ are evenly spaced in $[0,l]$, pick an integer $\lambda$ ranging from 1 to infinity, then set $\Delta x = \Delta x_0 / \lambda$. 
Then, given the initial bin function $B(x)$ replace it with a function $B_\lambda(x) \equiv \lambda B(\lambda x)$.
The limit of $\Delta x \to 0$ is then taken by letting $\lambda \to \infty$.
This process guarantees that the data remains finite and well-behaved in the limit.

To prove consistency, we ask if $T \to U$ in the limit of high resolution.
However as stated earlier, the two operators act on different spaces. 
Fortunately, in the Fourier representation, comparing the action of $T$ and $U$ is simple. 
Observing eqn.~(\ref{main}), one sees that the Fourier space representation of the difference operator, $T(k)$,  is defined for all $k < f_N$, whereas $U(k)$ is defined for all $k$.
As the spatial resolution increases, eventually the Nyquist frequency ($f_N = \pi /\Delta x$) will become greater than any fixed $k$.
Thus in the limit, the domain of definition of $T(k)$ approaches that of $U(k)$, and the two operators may be compared. 
We will show that $T(k) \to \bar{U}(k)$ as $\Delta x \to 0$, and since $\bar{U}(k) \to U(k)$ as $\Delta t \to 0$, this will prove that $T(k) \to U(k)$ in the joint limit. 
%
%
%

We can prove consistency in the translation-invariant case by adding some light restrictions: the response bins $B(x)$ are compactly supported on some strict subset $S$ of $[0,l]$, with bounded Fourier transform, and $\widehat{B}(0) \ne 0$. 
We also require that $U(k)\Phi(k) \to 0$ as $k \to \pm \infty$. 

The restrictions placed on the response bins are not too severe. 
The bounded Fourier transform requirement will almost always be true for any reasonable response bins. 
In particular, the \textit{Paley-Wiener theorem} \cite{Reedsimon} states that this requirement will hold for all smooth, compactly supported functions.  
The example box-response introduced in eqn.~(\ref{boxExample}) satisfies the restrictions, as will any smooth bump-function. 

We now seek a formula for $T(k)$ as a function of $\lambda$, which we call $T_\lambda(k)$. 
The compact support property of the bins allows us to exploit the fact that up to a  normalization constant, the coefficients $\widehat{B}(k)$ of the discrete values of $k$ in the Fourier series of the bins are the same as the values at $k$ in the continuous Fourier transform of $B(x)$: 
\begin{equation}
\int_0^l dx B(x) e^{-ikx} = \int_S dx B(x) e^{-ikx} = \int_{\mathbbm{R}} dx B(x) e^{-ikx}.
\end{equation}

This in turn allows us to exploit the scaling property of the Fourier transform to obtain a convenient form for $\widehat{B_\lambda} (k)$:
\begin{equation}
\widehat{B_\lambda} (k) = \lambda \widehat{B(\lambda x)} = \lambda \frac{1}{\lambda} \widehat{B}(k/\lambda) = \widehat{B}(k/\lambda). 
\end{equation} 

Now observe the sum over the Brillouin zones in eqn.~(\ref{main}). 
The formula contains a sum over  $b \in 2 f_N\mathbb{Z}$ where $f_N = \pi/\Delta x$ and thus $f_N^\lambda= \pi \lambda/\Delta x_0$  and $b^\lambda = 2 \pi n \lambda /\Delta x_0$ for $n \in \mathbb{Z}$.
We now insert the definitions of $\widehat{B_\lambda} (k)$ and $b^\lambda$ into eqn.~(\ref{main}), to yield a formula for $T_\lambda(k)$:

\begin{equation}\label{lambdaMain}
\frac{
 \sum_{n \in \mathbb{Z}} \bar{U}(k + \frac{2\pi n \lambda}{\Delta x_0}) \Phi(k + \frac{2\pi n \lambda}{\Delta x_0}) \big{|} \widehat{B}(\frac{1}{\lambda}(k + \frac{2\pi n \lambda}{\Delta x_0}))\big{|}^2
}
{
\sum_{m \in \mathbb{Z}} \Phi(k + \frac{2\pi m \lambda}{\Delta x_0}) \big{|} \widehat{B}(\frac{1}{\lambda}(k + \frac{2\pi m \lambda}{\Delta x_0}))\big{|}^2
}
\end{equation}


The $\lambda$ term inside $\widehat{B}$ can be absorbed to give:

\begin{equation}
| \widehat{B}(\frac{1}{\lambda}(k + \frac{2\pi n \lambda}{\Delta x_0}))|^2= | \widehat{B}(\frac{k}{\lambda} + \frac{2\pi n }{\Delta x_0}))|^2
\end{equation}
 
\noindent We expect that in the limit of $\lambda \to \infty$, the higher terms in the sums vanish, leaving only terms in the first Brillouin zone. 
That is to say, we can express the numerator of eqn.~(\ref{lambdaMain}) as:

\begin{align}\label{deltaExpansion}
&\bar{U}(k)\Phi(k)| \widehat{B}(\frac{k}{\lambda})|^2 + \\ 
&\underbrace{\sum_{n \ne 0} \bar{U}(k + \frac{2\pi n \lambda}{\Delta x_0}) \Phi(k + \frac{2\pi n \lambda}{\Delta x_0}) | \widehat{B}(\frac{k}{\lambda} + \frac{2\pi n }{\Delta x_0})|^2}_{\equiv \delta(k,\lambda) \to 0} \nonumber
\end{align}

Where we rename the sum $\delta(k,\lambda)$ to denote that the term (hopefully) vanishes as $\lambda \to \infty$. The denominator is expanded similarly and its vanishing term is denote by $\delta'(k,\lambda)$.
We wish to prove that the $\delta(k,\lambda)$ and $\delta'(k,\lambda)$ terms actually do vanish, as this would give us the desired result: 

\begin{align} \label{convergence}
& \lim_{\lambda \to \infty} T_\lambda(k) = \nonumber \\
 &\frac{\lim_{\lambda \to \infty} \sum_{n \in \mathbb{Z}} \bar{U}(k + \frac{2\pi n \lambda}{\Delta x_0}) \Phi(k + \frac{2\pi n \lambda}{\Delta x_0}) 
| \widehat{B}(\frac{k}{\lambda} + \frac{2\pi n }{\Delta x_0})|^2}
{\lim_{\lambda \to \infty}\sum_{m \in \mathbb{Z}} \Phi(k + \frac{2\pi m \lambda}{\Delta x_0}) | \widehat{B}(\frac{k}{\lambda} + \frac{2\pi m }{\Delta x_0})|^2} \nonumber \\
&= \frac{\lim_{\lambda \to \infty}\bar{U}(k) \Phi(k) | \widehat{B}(\frac{k}{\lambda} )|^2 +\delta(k,\lambda)}{\lim_{\lambda \to \infty} \Phi(k) | \widehat{B}(\frac{k}{\lambda} )|^2 +\delta'(k,\lambda)} \nonumber \\
& = \frac{\bar{U}(k) \Phi(k) | \widehat{B}(0)|^2}{ \Phi(k) | \widehat{B}(0)|^2} = \bar{U}(k).
\end{align}

This works provided $\widehat{B}(0) \ne 0$, so that the above denominator remains nonzero, and the equation remains well-defined. 
For $\delta$ and $\delta'$, each individual term in the sum (which we denote by $\delta_n(k, \lambda)$) approaches zero in the limit of $\lambda \to \infty$, because $\bar{U}(k)\Phi(k)$ and $\Phi(k)$ go to zero at large $|k|$, by assumption. 
Therefore we want to swap the limit and the infinite sum. 

Elementary functional analysis states that this is possible if and only if the sequence of $\delta_n$-terms converges uniformly to zero in $n$.
We remind the reader that a sequence of functions $\delta_n$ converges uniformly to zero if for any positive $\epsilon$, 
there is an $N$ such that $\forall n \ge N$, $|\delta_n(\lambda)| < \epsilon$ for all values of $\lambda$. 

We prove uniform convergence for $\delta$, and $\delta'$ follows trivially.
We consider the positive-$n$ half of the sum first, and the negative-$n$ half will also follow trivially. 
In this case,
\begin{align}
 \delta_n(\lambda) = &\bar{U}(k + \frac{2\pi n \lambda}{\Delta x_0}) \Phi(k + \frac{2\pi n \lambda}{\Delta x_0}) 
| \widehat{B}(\frac{k}{\lambda} + \frac{2\pi n }{\Delta x_0})|^2 \nonumber \\
& \le |\bar{U}(k + \frac{2\pi n \lambda}{\Delta x_0}) \Phi(k + \frac{2\pi n \lambda}{\Delta x_0})|C,
\end{align}
where we bounded the function $|\widehat{B}(\frac{k}{\lambda} + \frac{2\pi n }{\Delta x_0})|^2 < C$
for some constant $C$, which we may do by assumption\footnote{
To understand why this bound is necessary, notice that the bin terms do not vanish in the limit of large $\lambda$.
Intuitively, this is because as the bins become narrower, their Fourier transforms widen out, at the exact same rate as the Nyquist frequency is increasing.
}. 

We now use the condition $\Phi(k)\bar{U}(k)\to 0$ as $k \to \infty$ to show convergence. 
This condition means that for $|k|$ large enough $\bar{U}(k)\Phi(k)$ can be bounded by some monotonically decreasing function of $|k|$, call it $g(|k|)$. 
We start by finding a bound for 
 $\lambda=1$, and then show that this bound holds for all $\lambda$. 
For $\lambda=1$, and the desired $\epsilon$ bound, we can pick some $n$ large enough such that we are in this decreasing regime, hence:

\begin{equation}
\delta_n \le |\bar{U}(k + \frac{2\pi n \lambda}{\Delta x_0}) \Phi(k + \frac{2\pi n \lambda}{\Delta x_0})|C < g(k + \frac{2\pi n \lambda}{\Delta x_0}) < \epsilon.
\end{equation}
For higher $\lambda$ and large $n$,  $|k + \frac{2\pi n}{\Delta x_0}| < |k + \frac{2\pi n \lambda}{\Delta x_0}|$, and since we have taken $n$ to be large enough that we are in the decreasing regime, the $g(k)$ bound also holds. 
Thus the bound holds for all $\lambda$. 
The sequence of functions is therefore uniformly convergent, and eqn.~(\ref{convergence}) holds. 
We can now state:

\begin{theorem}\label{IFDconsistency}
For a 1-D translationally-invariant IFD scheme, whose response bins $B(x)$ are compactly supported on a strict subset $S \subset [0,l]$ with bounded Fourier transform and $\widehat{B}(0) \ne 0$, and some time-order approximation $\bar{U}(k)$ to $U(k)$, then the scheme is consistent provided $\lim_{k \to \infty} \bar{U}(k)\Phi(k) =0$. 
\end{theorem}

Important to note is that we only require $\bar{U}(k)\Phi(k) \to 0$, not $U(k)\Phi(k) \to 0$. 
For derivative operators s.t. 
$U = \exp(\Delta t \partial_x) =\exp(i\Delta t k)$ or similar, this would require that the prior covariance, $\Phi(x)$, is infinitely differentiable, a.k.a smooth. 
Using the approximated time expansion, the prior covariance only needs to be as many-times differentiable as the order of the expansion dictates. 
%

%
%
%
%
%
%
%
%

\subsection{Error scaling}
We seek an estimate of the one-step error $E$ by calculating the difference in the operator norm:

\begin{equation}
E \propto \|T(\Delta t, \Delta x) - U(\Delta t)\|
\end{equation}

\noindent and analyzing the rate of convergence in terms of $\mathcal{O}(\Delta x)$ and $\mathcal{O}(\Delta t)$. 
We calculate the error for a fixed value of $k$, thus $E=E(k)$, although it will be seen shortly that the scaling of the error (which is the quantity of interest) is independent of $k$. 
We repeat the same construction as before by scaling the resolution with $\lambda$. We insert the difference operator in eqn.~(\ref{lambdaMain}) into the error definition, yielding:

\begin{align} \label{fourier_error}
 &E(k) =  \\
& \bigg{|}\frac{ \sum_{n \in \mathbb{Z}} \bar{U}(k + \frac{2\pi n \lambda}{\Delta x_0}) \Phi(k + \frac{2\pi n \lambda}{\Delta x_0}) 
| \widehat{B}(\frac{k}{\lambda} + \frac{2\pi n }{\Delta x_0})|^2}
{\sum_{m \in \mathbb{Z}} \Phi(k + \frac{2\pi m \lambda}{\Delta x_0}) | \widehat{B}(\frac{k}{\lambda} + \frac{2\pi m }{\Delta x_0})|^2}- U(k)\bigg{|}. \nonumber
\end{align}

We use the expansion $\bar{U}=\sum_{p=0}^{\alpha}(\Delta t L)^p/p!$ to find the error in terms of powers of $L$.

\begin{align}\label{errorExpansion}
 &E(k) \le \sum_{p=0}^{\alpha}\frac{\Delta t^p}{p!} \times \\ 
&\bigg{|}\frac{ \sum_{n \in \mathbb{Z}} L^p(k + \frac{2\pi n \lambda}{\Delta x_0}) \Phi(k + \frac{2\pi n \lambda}{\Delta x_0}) 
| \widehat{B}(\frac{k}{\lambda} + \frac{2\pi n }{\Delta x_0})|^2}
{\sum_{m \in \mathbb{Z}} \Phi(k + \frac{2\pi m \lambda}{\Delta x_0}) | \widehat{B}(\frac{k}{\lambda} + \frac{2\pi m }{\Delta x_0})|^2}- L^p(k)\bigg{|}. \nonumber
\end{align}

For each of the $\Delta t^p$ terms, we will analyze the scaling of the fraction inside the absolute value, and then later find an estimate of the total error scaling. 

In the limit of high resolutions, we can expand the numerator of said fraction in the same way that we did in eqn.~(\ref{deltaExpansion}): $L^p(k)\Phi(k)| \widehat{B}(\frac{k}{\lambda})|^2 +\epsilon(k,\lambda)$ for some function $\epsilon$, which goes to zero as $\lambda \to \infty$. 
We expand the denominator as $\Phi(k)| \widehat{B}(\frac{k}{\lambda})|^2 +\delta(k,\lambda)$, with $\delta$ being some other small vanishing function. 

The strategy is then to find an expression for the fraction in terms of $\epsilon$ and $\delta$, then bound each term and analyze how fast they approach zero.
We use the Taylor expansion for $1/(1-x) \approx 1+x+ x^2+\cdots$ to expand the denominator in eqn.~(\ref{errorExpansion}) into:

\begin{align}
&\frac{1}{\Phi(k)| \widehat{B}(\frac{k}{\lambda})|^2 +\delta(k,\lambda)}= \\
&\frac{1}{\Phi(k)| \widehat{B}(\frac{k}{\lambda})|^2} - \frac{\delta(k,\lambda)}{(\Phi(k)| \widehat{B}(\frac{k}{\lambda})|^2)^2} + \cdots. \nonumber
\end{align}

We then multiply the numerator by the denominator, which gives:

\begin{align}\label{EpsDelta}
&\bigg{(}L^p(k)\Phi(k)| \widehat{B}(\frac{k}{\lambda})|^2 + \epsilon(k,\lambda) \bigg{)} \times \\
&\bigg{(} \frac{1}{\Phi(k)| \widehat{B}(\frac{k}{\lambda})|^2} - \frac{\delta(k,\lambda)}{(\Phi(k)| \widehat{B}(\frac{k}{\lambda})|^2)^2} + \cdots \bigg{)} \nonumber \\
&=L^p(k) + \frac{\epsilon(k,\lambda)}{\Phi(k)| \widehat{B}(\frac{k}{\lambda})|^2} - \frac{L^p(k)\delta(k,\lambda)}{\Phi(k)| \widehat{B}(\frac{k}{\lambda})|^2}  + \cdots. \nonumber
\end{align}

Calculating the power-law scaling of the above terms is complicated by the fact that each has a $|\widehat{B}(\frac{k}{\lambda})|^2$ in the denominator, which has its own scaling w.r.t. $\lambda$. 
Exploiting the property $\widehat{B}(0) \ne 0$, allows us to write out each of these as a Taylor series, and then reuse the $1/(1-x)$ expansion: 

\begin{align}
\frac{1}{\Phi(k)| \widehat{B}(\frac{k}{\lambda})|^2} & = \frac{1}{\Phi(k)| \widehat{B}(0)|^2 + \mathcal{O}(1/\lambda) +\cdots} \nonumber \\
&= \frac{1}{\Phi(k)| \widehat{B}(0)|^2} + \mathcal{O}(1/\lambda) +\cdots
\label{eq:}
\end{align}

We then see that however fast $\epsilon(k,\lambda)$ goes to zero, $\epsilon(k,\lambda) \mathcal{O}(1/\lambda)$ goes to zero faster. 
Since only the slowest-converging terms are of interest, we can replace $\frac{1}{\Phi(k)| \widehat{B}(\frac{k}{\lambda})|^2}$ with $\frac{1}{\Phi(k)| \widehat{B}(0)|^2}$ without any adverse effects. 

The scaling of the $\epsilon$ and $\delta$ terms can only be estimated if the scaling behaviour of the prior and $L(k)$ are known. 
To this end, suppose that as $|k|$ becomes large, $\Phi(k)$ can be bounded by some decreasing power law in $k$, $|k|^{-\beta}$ for $\beta$ positive. 
We also assume that $L(k)$ can be bounded by some $|k|^\gamma$ for $\gamma$ positive, as $L$ will typically be a derivative operator, with $\partial_x^n = (ik)^n$. 
Then $L^p(k)$ will be bounded by $|k|^{p\gamma}$.
There will be constants of proportionality, but they are irrelevant with respect to the scaling. 

Using the uniform bound $C$ from before, we can bound the $\epsilon$ term by:
\begin{align}
|\epsilon(k,\lambda)| & = \bigg{|}\sum_{n \ne 0} L^p(k + \frac{2\pi n \lambda}{\Delta x_0}) \Phi(k + \frac{2\pi n \lambda}{\Delta x_0}) | \widehat{B}(\frac{k}{\lambda} + \frac{2\pi n }{\Delta x_0})|^2 
\bigg{|}\nonumber \\
&\le C^2 \sum_{n \ne 0}\bigg{|}  L^p(k + \frac{2\pi n \lambda}{\Delta x_0}) \Phi(k + \frac{2\pi n \lambda}{\Delta x_0})   \bigg{|}\nonumber \\
&\le C^2\sum_{n \ne 0}\bigg{|}  \frac{2\pi n \lambda}{\Delta x_0}   \bigg{|}^{p\gamma - \beta} = \lambda^{p\gamma - \beta}C^2\sum_{n \ne 0}\bigg{|}  \frac{2\pi n}{\Delta x_0}   \bigg{|}^{p\gamma - \beta}.
\end{align}

The term inside the sum is independent of $\lambda$. 
Therefore, this bound scales as $\mathcal{O}(\lambda^{p\gamma-\beta})$, which we identify with $\mathcal{O}(\Delta x^{\beta-p\gamma})$, since $\Delta x = \Delta x_0/\lambda$. 
We repeat the argument with $\delta$, and obtain a term of order $\mathcal{O}(\Delta x^{\beta})$. 
Thus eqn.~(\ref{EpsDelta}) scales as:
\begin{align}
&L^p(k) + \mathcal{O}(\Delta x^{\beta})+ \mathcal{O}(\Delta x^{\beta-p\gamma}) \nonumber \\
& =L^p(k)+ \boxed{ \mathcal{O}(\Delta x^{\beta-p\gamma})},
\end{align}
 
\noindent yielding a total time and space error scaling of $\mathcal{O}(\Delta t^p) \mathcal{O}(\Delta x^{\beta-p\gamma})$. 
The other $\mathcal{O}$ term vanishes because only the term with the worst scaling (lowest power) contributes. 
The total error scaling in eqn.~(\ref{errorscaling}) is determined by the sum of the individual $p$ terms:

\begin{equation}\label{errorscaling}
\boxed{E \propto \sum_{p=0}^{\alpha}\mathcal{O}(\Delta t^p \Delta x^{\beta-p\gamma})}, 
\end{equation}

although the error will be bounded by the worst scaling of any of the individual terms. 
 We see from this formula that taking higher orders in $\Delta t$ decreases the spatial order. 
This is fine for $L =\partial_x$, because the total order remains the same, but for higher derivatives, the spatial order decreases faster in $p$ than the time order increases. 
If $\Delta x$ and $\Delta t \to 0$ at a proportional rate, this will decrease the total order and making the overall error scaling worse. 

This can be thought of in the following way: if the prior covariance only drops off as some power $\beta$, then it is only $\beta$ times differentiable, so it is not smooth. 
Taking higher orders in the expansion $\bar{U}=\sum_{p=0}^{\alpha}(\Delta t L)^p/p!$ involves taking derivatives of ever-higher order, and thus at some point the $L \Phi(x)$ term in the difference operator can no longer be calculated. 
The bin functions do not appear in the above expression, because in the limit of high resolutions, they tend to approximate delta functions, and their exact form becomes irrelevant.

The consequences of this formula deserve some thought, particularly the troubling implication that going to higher orders in time can in fact decrease the quality of the simulation. 
First, it should be noted that higher-order schemes are not necessarily better, depending on the task. 
For example, according to the \textit{Godunov Theorem} \cite[p.~280]{Godunov}, higher order schemes have a tendency to develop spurious oscillations around shocks. 
It should also be noted that the above formula applies in the high resolution (and thus high-$k$) limit. 
One could conceivably introduce a prior covariance which has a cutoff at high $k$, or perhaps one whose value drops of exponentially with $k$. 
An exponentially-falling prior covariance would then raise the prospect of a finite-difference scheme with intermediate error scaling, however the implications of such a scheme are not yet clear.  

\subsection{Generalization to higher dimensions}
The previous derivation was only presented for the one-dimensional case for the sake of pedagogical clarity.
If we extend to the $M$-dimensional case, then $x$ and $k$ become vectors $\vec{x}$, $\vec{k}$, and the simulation domain becomes $\prod_i^M[0,l_i]$.
Eqn.~(\ref{main}) becomes a sum over vectors $\vec{k}$ less than $\vec{f_N}$ where the Nyquist frequency is now a vector due to the (possibly) differing resolutions along each grid direction, and likewise the sum over Brillouin zones is also vector-valued.

To prove consistency, the assumptions do not need to be tightened, except that we must now specify $U(\vec{k})\Phi(\vec{k}) \to 0$ as $\| \vec{k}\| \to \infty$.
We also need the resolution to be increased in all spatial dimensions at the same rate, so $\Delta x $ becomes $\Delta \vec{x} = \Delta \vec{x_0} / \lambda$.
The proof then proceeds as before.

In order to show that the error scaling formula~(\ref{errorscaling}) holds in higher dimensions, one needs to put a new bound on $L(\vec{k})$ such that it is bounded by $\| \vec{k}\| ^\gamma$.
This property is easily fulfilled by many differential operators, such as $L = \partial_x^2 + \partial_y^2 + \partial_z^2$ for example. 
Likewise, we assume that $\Phi(k)$ can now be bounded by some $\| \vec{k}\|^\beta$, and the proof proceeds as before. 

\section{Conclusions}

We have now proved consistency, and found an estimate of the error scaling for IFD schemes, using a set of strong simplifying assumptions, which we grouped together under the name of a \textit{translation-invariant scheme}. 
These assumptions were:

\begin{itemize}
\item ``Linear case'' of IFD: linear differential equation, linear measurements with additive noise, and Gaussian prior distribution of the fields.
\item Translation and time invariance of all the above quantities.
\item Box-shaped simulation space with periodic boundary conditions.
\item A response $R$ which integrates the field $\phi$ against an evenly-spaced grid of bin functions. 
\item A bin functions $B(x)$ which is compactly supported on a strict subset of the simulation space, has bounded Fourier transform, and whose Fourier transform $\widehat{B}(\vec{k})$ has $\widehat{B}(0) \ne 0$.
\item $\bar{U}(\vec{k})\Phi(\vec{k}) \to 0$ as $\|k\| \to \infty$.
\end{itemize}

\noindent To obtain an estimate of the error scaling, we needed to assume: 
\begin{itemize}
\item The operators $L(\vec{k})$ and $\Phi(\vec{k})$ may be bounded by power-laws $\|k\|^\gamma$ and  $\|k\|^{-\beta}$ for $\beta, \gamma > 0$ respectively, at large values of $\|k\|$.
\end{itemize}
These restrictions mean that the results in this paper are only directly applicable to a very small subset of the simulation schemes that may be constructed using IFD. 
Given the immense amount of freedom inherent in the IFD framework, it is doubtful that a general analytic proof of consistency will be achievable. 
This paper should be instead taken as a general indication that IFD is at least a sensible methodology. 

That being said, it is expected that the above assumptions could be weakened in order to obtain a stronger result. 
Most obviously, the fact that the difference operator can be expressed using a \textit{sum over Brillouin zones} immediately suggests that these results could be extended to a simulation over any periodic lattice of data points; not just rectangular domains. 

The restrictions on the bin functions are relatively weak. 
The compact support requirement simply ensures that the response corresponds to some sort of local measurement of the field. 
The requirement that $\widehat{B}(0) \ne 0$ deserves some discussion however. 
This requirement, rather than being physically motivated,  was inserted solely to avoid the occurrence of $0/0$ terms in the limit of high resolutions. 
It may however reflect a physical requirement. 
Take, for example, a bin function $B(x)$ which is everywhere positive, and is symmetric and peaked about zero. 
It will satisfy $\widehat{B}(0) \ne 0$, and in the limit of high resolutions, this bin will approach a delta function, and will represent a sample of the field value at that point. 
In contrast, take $xB(x)$; this function is now odd, and in the limit of high resolutions, this will approach something that represents a point sample of the \textit{derivative} of the field about that point. 
Attempting to apply IFD to reconstructions of the derivative of the field may give nonsensical results, which is what the $\widehat{B}(0) \ne 0$ requirement may be implying. 

Removing the translation-invariance requirements would be extremely desirable, but much more difficult. 
The main reason the Fourier approach was necessary was to allow the inversion of the $(R \Phi R^\dagger)_{ij}$ matrices. 
With a reasonable smoothness prior, these matrices tend to be relatively local in the spatial indices. 
However, inversion is a nonlocal problem, which makes the inverses of these matrices dependent on the global geometry of the simulation domain, and makes them very difficult to study analytically. 
The use of periodic boundary conditions allowed us to sidestep this consideration. 
Any proof seeking to show consistency and convergence in the non-translationally-invariant case would probably have to use a different approach to what we have done here. 

Finally, an extension of these results outside of the linear regime is self-evidently desirable, but may present some significant challenges. 
In particular, IFT inference problems including nonlinear responses and non-Gaussian priors on the fields typically result in a need to calculate Feynmann diagrams. 
Integrating these into an analytic proof of consistency will be challenging to say the least. 
\bibliography{mybib}

\begin{thebibliography}{22}%
\makeatletter
\providecommand \@ifxundefined [1]{%
 \@ifx{#1\undefined}
}%
\providecommand \@ifnum [1]{%
 \ifnum #1\expandafter \@firstoftwo
 \else \expandafter \@secondoftwo
 \fi
}%
\providecommand \@ifx [1]{%
 \ifx #1\expandafter \@firstoftwo
 \else \expandafter \@secondoftwo
 \fi
}%
\providecommand \natexlab [1]{#1}%
\providecommand \enquote  [1]{``#1''}%
\providecommand \bibnamefont  [1]{#1}%
\providecommand \bibfnamefont [1]{#1}%
\providecommand \citenamefont [1]{#1}%
\providecommand \href@noop [0]{\@secondoftwo}%
\providecommand \href [0]{\begingroup \@sanitize@url \@href}%
\providecommand \@href[1]{\@@startlink{#1}\@@href}%
\providecommand \@@href[1]{\endgroup#1\@@endlink}%
\providecommand \@sanitize@url [0]{\catcode `\\12\catcode `\$12\catcode
  `\&12\catcode `\#12\catcode `\^12\catcode `\_12\catcode `\%12\relax}%
\providecommand \@@startlink[1]{}%
\providecommand \@@endlink[0]{}%
\providecommand \url  [0]{\begingroup\@sanitize@url \@url }%
\providecommand \@url [1]{\endgroup\@href {#1}{\urlprefix }}%
\providecommand \urlprefix  [0]{URL }%
\providecommand \Eprint [0]{\href }%
\providecommand \doibase [0]{http://dx.doi.org/}%
\providecommand \selectlanguage [0]{\@gobble}%
\providecommand \bibinfo  [0]{\@secondoftwo}%
\providecommand \bibfield  [0]{\@secondoftwo}%
\providecommand \translation [1]{[#1]}%
\providecommand \BibitemOpen [0]{}%
\providecommand \bibitemStop [0]{}%
\providecommand \bibitemNoStop [0]{.\EOS\space}%
\providecommand \EOS [0]{\spacefactor3000\relax}%
\providecommand \BibitemShut  [1]{\csname bibitem#1\endcsname}%
\let\auto@bib@innerbib\@empty
\bibitem [{\citenamefont {En\ss{}lin}(2013)}]{IFD}%
  \BibitemOpen
  \bibfield  {author} {\bibinfo {author} {\bibfnamefont {T.~A.}\ \bibnamefont
  {En\ss{}lin}},\ }\bibfield  {title} {\enquote {\bibinfo {title} {Information
  field dynamics for simulation scheme construction},}\ }\href {\doibase
  10.1103/PhysRevE.87.013308} {\bibfield  {journal} {\bibinfo  {journal} {Phys.
  Rev. E}\ }\textbf {\bibinfo {volume} {87}},\ \bibinfo {pages} {013308}
  (\bibinfo {year} {2013})}\BibitemShut {NoStop}%
\bibitem [{\citenamefont {{M{\"u}nch}}(2014)}]{IFDMath}%
  \BibitemOpen
  \bibfield  {author} {\bibinfo {author} {\bibfnamefont {C.}~\bibnamefont
  {{M{\"u}nch}}},\ }\emph {\bibinfo {title} {{Mathematical foundation of
  Information Field Dynamics (revised version)}}},\ \href@noop {} {Master's
  thesis} (\bibinfo {year} {2014}),\ \Eprint {http://arxiv.org/abs/1412.1226}
  {arXiv:1412.1226 [math.DS]} \BibitemShut {NoStop}%
\bibitem [{\citenamefont {{Leike}}\ and\ \citenamefont
  {{En{\ss}lin}}(2017)}]{ReimarTowards}%
  \BibitemOpen
  \bibfield  {author} {\bibinfo {author} {\bibfnamefont {R.~H.}\ \bibnamefont
  {{Leike}}}\ and\ \bibinfo {author} {\bibfnamefont {T.~A.}\ \bibnamefont
  {{En{\ss}lin}}},\ }\bibfield  {title} {\enquote {\bibinfo {title} {{Towards
  information optimal simulation of partial differential equations}},}\
  }\href@noop {} {\bibfield  {journal} {\bibinfo  {journal} {ArXiv e-prints}\ }
  (\bibinfo {year} {2017})},\ \Eprint {http://arxiv.org/abs/1709.02859}
  {arXiv:1709.02859 [stat.ME]} \BibitemShut {NoStop}%
\bibitem [{\citenamefont {Hennig}\ \emph {et~al.}(2015)\citenamefont {Hennig},
  \citenamefont {Osborne},\ and\ \citenamefont {Girolami}}]{HenOsbGirRSPA2015}%
  \BibitemOpen
  \bibfield  {author} {\bibinfo {author} {\bibfnamefont {P.}~\bibnamefont
  {Hennig}}, \bibinfo {author} {\bibfnamefont {M.A.}\ \bibnamefont {Osborne}},
  \ and\ \bibinfo {author} {\bibfnamefont {M.}~\bibnamefont {Girolami}},\
  }\bibfield  {title} {\enquote {\bibinfo {title} {Probabilistic numerics and
  uncertainty in computations},}\ }\href@noop {} {\bibfield  {journal}
  {\bibinfo  {journal} {Proceedings of the Royal Society of London A:
  Mathematical, Physical and Engineering Sciences}\ }\textbf {\bibinfo {volume}
  {471}} (\bibinfo {year} {2015})}\BibitemShut {NoStop}%
\bibitem [{\citenamefont {Diaconis}(1988)}]{Diaconis}%
  \BibitemOpen
  \bibfield  {author} {\bibinfo {author} {\bibfnamefont {P.}~\bibnamefont
  {Diaconis}},\ }\bibfield  {title} {\enquote {\bibinfo {title} {Bayesian
  numerical analysis},}\ }\href@noop {} {\bibfield  {journal} {\bibinfo
  {journal} {Statistical decision theory and related topics IV}\ }\textbf
  {\bibinfo {volume} {1}},\ \bibinfo {pages} {163--175} (\bibinfo {year}
  {1988})}\BibitemShut {NoStop}%
\bibitem [{\citenamefont {O’Hagan}(1992)}]{oHagan}%
  \BibitemOpen
  \bibfield  {author} {\bibinfo {author} {\bibfnamefont {A.}~\bibnamefont
  {O’Hagan}},\ }\bibfield  {title} {\enquote {\bibinfo {title} {Some bayesian
  numerical analysis},}\ }\href@noop {} {\bibfield  {journal} {\bibinfo
  {journal} {Bayesian Statistics}\ }\textbf {\bibinfo {volume} {4}},\ \bibinfo
  {pages} {4--2} (\bibinfo {year} {1992})}\BibitemShut {NoStop}%
\bibitem [{\citenamefont {Larkin}(1972)}]{Larkin}%
  \BibitemOpen
  \bibfield  {author} {\bibinfo {author} {\bibfnamefont {F.~M.}\ \bibnamefont
  {Larkin}},\ }\bibfield  {title} {\enquote {\bibinfo {title} {Gaussian measure
  in hilbert space and applications in numerical analysis},}\ }\href
  {http://www.jstor.org/stable/44236270} {\bibfield  {journal} {\bibinfo
  {journal} {The Rocky Mountain Journal of Mathematics}\ }\textbf {\bibinfo
  {volume} {2}},\ \bibinfo {pages} {379--421} (\bibinfo {year}
  {1972})}\BibitemShut {NoStop}%
\bibitem [{\citenamefont {Skilling}(1991)}]{Skilling1}%
  \BibitemOpen
  \bibfield  {author} {\bibinfo {author} {\bibfnamefont {J.}~\bibnamefont
  {Skilling}},\ }\bibfield  {title} {\enquote {\bibinfo {title} {{Bayesian
  solution of ordinary differential equations}},}\ }\href@noop {} {\bibfield
  {journal} {\bibinfo  {journal} {Maximum Entropy and Bayesian Methods,
  Seattle}\ } (\bibinfo {year} {1991})}\BibitemShut {NoStop}%
\bibitem [{\citenamefont {Coulibaly}\ and\ \citenamefont
  {Lécot}(1999)}]{Lecot}%
  \BibitemOpen
  \bibfield  {author} {\bibinfo {author} {\bibfnamefont {I.}~\bibnamefont
  {Coulibaly}}\ and\ \bibinfo {author} {\bibfnamefont {C.}~\bibnamefont
  {Lécot}},\ }\bibfield  {title} {\enquote {\bibinfo {title} {C.: A
  quasi-randomized runge-kutta method},}\ }\href@noop {} {\bibfield  {journal}
  {\bibinfo  {journal} {Math. Comput.}\ }\textbf {\bibinfo {volume} {68}},\
  \bibinfo {pages} {651--659} (\bibinfo {year} {1999})}\BibitemShut {NoStop}%
\bibitem [{\citenamefont {Stengle}(1995)}]{Stengle}%
  \BibitemOpen
  \bibfield  {author} {\bibinfo {author} {\bibfnamefont {G.}~\bibnamefont
  {Stengle}},\ }\bibfield  {title} {\enquote {\bibinfo {title} {Error analysis
  of a randomized numerical method},}\ }\href {\doibase 10.1007/s002110050113}
  {\bibfield  {journal} {\bibinfo  {journal} {Numerische Mathematik}\ }\textbf
  {\bibinfo {volume} {70}},\ \bibinfo {pages} {119--128} (\bibinfo {year}
  {1995})}\BibitemShut {NoStop}%
\bibitem [{\citenamefont {{Briol}}\ \emph {et~al.}(2015)\citenamefont
  {{Briol}}, \citenamefont {{Oates}}, \citenamefont {{Girolami}}, \citenamefont
  {{Osborne}},\ and\ \citenamefont {{Sejdinovic}}}]{Briol}%
  \BibitemOpen
  \bibfield  {author} {\bibinfo {author} {\bibfnamefont {F.-X.}\ \bibnamefont
  {{Briol}}}, \bibinfo {author} {\bibfnamefont {C.~J.}\ \bibnamefont
  {{Oates}}}, \bibinfo {author} {\bibfnamefont {M.}~\bibnamefont {{Girolami}}},
  \bibinfo {author} {\bibfnamefont {M.~A.}\ \bibnamefont {{Osborne}}}, \ and\
  \bibinfo {author} {\bibfnamefont {D.}~\bibnamefont {{Sejdinovic}}},\
  }\bibfield  {title} {\enquote {\bibinfo {title} {{Probabilistic Integration:
  A Role in Statistical Computation?}}}\ }\href@noop {} {\bibfield  {journal}
  {\bibinfo  {journal} {ArXiv e-prints}\ } (\bibinfo {year} {2015})},\ \Eprint
  {http://arxiv.org/abs/1512.00933} {arXiv:1512.00933 [stat.ML]} \BibitemShut
  {NoStop}%
\bibitem [{\citenamefont {Schober}\ \emph {et~al.}(2014)\citenamefont
  {Schober}, \citenamefont {Duvenaud},\ and\ \citenamefont {Hennig}}]{Schober}%
  \BibitemOpen
  \bibfield  {author} {\bibinfo {author} {\bibfnamefont {M.}~\bibnamefont
  {Schober}}, \bibinfo {author} {\bibfnamefont {D.K.}\ \bibnamefont
  {Duvenaud}}, \ and\ \bibinfo {author} {\bibfnamefont {P.}~\bibnamefont
  {Hennig}},\ }\bibfield  {title} {\enquote {\bibinfo {title} {Probabilistic
  ode solvers with runge-kutta means},}\ }in\ \href
  {http://papers.nips.cc/paper/5451-probabilistic-ode-solvers-with-runge-kutta-means.pdf}
  {\emph {\bibinfo {booktitle} {Advances in Neural Information Processing
  Systems 27}}},\ \bibinfo {editor} {edited by\ \bibinfo {editor}
  {\bibfnamefont {Z.}~\bibnamefont {Ghahramani}}, \bibinfo {editor}
  {\bibfnamefont {M.}~\bibnamefont {Welling}}, \bibinfo {editor} {\bibfnamefont
  {C.}~\bibnamefont {Cortes}}, \bibinfo {editor} {\bibfnamefont {N.~D.}\
  \bibnamefont {Lawrence}}, \ and\ \bibinfo {editor} {\bibfnamefont {K.~Q.}\
  \bibnamefont {Weinberger}}}\ (\bibinfo  {publisher} {Curran Associates,
  Inc.},\ \bibinfo {year} {2014})\ pp.\ \bibinfo {pages} {739--747}\BibitemShut
  {NoStop}%
\bibitem [{\citenamefont {{Conrad}}\ \emph {et~al.}(2015)\citenamefont
  {{Conrad}}, \citenamefont {{Girolami}}, \citenamefont {{S{\"a}rkk{\"a}}},
  \citenamefont {{Stuart}},\ and\ \citenamefont {{Zygalakis}}}]{Conrad}%
  \BibitemOpen
  \bibfield  {author} {\bibinfo {author} {\bibfnamefont {P.~R.}\ \bibnamefont
  {{Conrad}}}, \bibinfo {author} {\bibfnamefont {M.}~\bibnamefont
  {{Girolami}}}, \bibinfo {author} {\bibfnamefont {S.}~\bibnamefont
  {{S{\"a}rkk{\"a}}}}, \bibinfo {author} {\bibfnamefont {A.}~\bibnamefont
  {{Stuart}}}, \ and\ \bibinfo {author} {\bibfnamefont {K.}~\bibnamefont
  {{Zygalakis}}},\ }\bibfield  {title} {\enquote {\bibinfo {title}
  {{Probability Measures for Numerical Solutions of Differential Equations}},}\
  }\href@noop {} {\bibfield  {journal} {\bibinfo  {journal} {ArXiv e-prints}\ }
  (\bibinfo {year} {2015})},\ \Eprint {http://arxiv.org/abs/1506.04592}
  {arXiv:1506.04592 [stat.ME]} \BibitemShut {NoStop}%
\bibitem [{\citenamefont {Archambeau}\ \emph {et~al.}(2007)\citenamefont
  {Archambeau}, \citenamefont {Cornford}, \citenamefont {Opper},\ and\
  \citenamefont {Shawe-Taylor}}]{Archambeau}%
  \BibitemOpen
  \bibfield  {author} {\bibinfo {author} {\bibfnamefont {C.}~\bibnamefont
  {Archambeau}}, \bibinfo {author} {\bibfnamefont {D.}~\bibnamefont
  {Cornford}}, \bibinfo {author} {\bibfnamefont {M.}~\bibnamefont {Opper}}, \
  and\ \bibinfo {author} {\bibfnamefont {J.}~\bibnamefont {Shawe-Taylor}},\
  }\bibfield  {title} {\enquote {\bibinfo {title} {Gaussian process
  approximations of stochastic differential equations},}\ }in\ \href
  {http://proceedings.mlr.press/v1/archambeau07a.html} {\emph {\bibinfo
  {booktitle} {Gaussian Processes in Practice}}},\ \bibinfo {series}
  {Proceedings of Machine Learning Research}, Vol.~\bibinfo {volume} {1},\
  \bibinfo {editor} {edited by\ \bibinfo {editor} {\bibfnamefont {Neil~D.}\
  \bibnamefont {Lawrence}}, \bibinfo {editor} {\bibfnamefont {Anton}\
  \bibnamefont {Schwaighofer}}, \ and\ \bibinfo {editor} {\bibfnamefont
  {Joaquin~Quiñonero}\ \bibnamefont {Candela}}}\ (\bibinfo  {publisher}
  {PMLR},\ \bibinfo {address} {Bletchley Park, UK},\ \bibinfo {year} {2007})\
  pp.\ \bibinfo {pages} {1--16}\BibitemShut {NoStop}%
\bibitem [{\citenamefont {Leike}\ and\ \citenamefont {Enßlin}(2017)}]{Reimar}%
  \BibitemOpen
  \bibfield  {author} {\bibinfo {author} {\bibfnamefont {R.~H.}\ \bibnamefont
  {Leike}}\ and\ \bibinfo {author} {\bibfnamefont {T.~A.}\ \bibnamefont
  {Enßlin}},\ }\bibfield  {title} {\enquote {\bibinfo {title} {Optimal belief
  approximation},}\ }\href {http://www.mdpi.com/1099-4300/19/8/402} {\bibfield
  {journal} {\bibinfo  {journal} {Entropy}\ }\textbf {\bibinfo {volume} {19}}
  (\bibinfo {year} {2017})}\BibitemShut {NoStop}%
\bibitem [{\citenamefont {{En{\ss}lin}}(2018)}]{2018arXiv180403350E}%
  \BibitemOpen
  \bibfield  {author} {\bibinfo {author} {\bibfnamefont {T.~A.}\ \bibnamefont
  {{En{\ss}lin}}},\ }\bibfield  {title} {\enquote {\bibinfo {title}
  {{Information theory for fields}},}\ }\href@noop {} {\bibfield  {journal}
  {\bibinfo  {journal} {ArXiv e-prints}\ } (\bibinfo {year} {2018})},\ \Eprint
  {http://arxiv.org/abs/1804.03350} {arXiv:1804.03350} \BibitemShut {NoStop}%
\bibitem [{\citenamefont {En\ss{}lin}\ \emph {et~al.}(2009)\citenamefont
  {En\ss{}lin}, \citenamefont {Frommert},\ and\ \citenamefont {Kitaura}}]{IFT}%
  \BibitemOpen
  \bibfield  {author} {\bibinfo {author} {\bibfnamefont {T.A}\ \bibnamefont
  {En\ss{}lin}}, \bibinfo {author} {\bibfnamefont {M.}~\bibnamefont
  {Frommert}}, \ and\ \bibinfo {author} {\bibfnamefont {F.S}\ \bibnamefont
  {Kitaura}},\ }\bibfield  {title} {\enquote {\bibinfo {title} {Information
  field theory for cosmological perturbation reconstruction and nonlinear
  signal analysis},}\ }\href {\doibase 10.1103/PhysRevD.80.105005} {\bibfield
  {journal} {\bibinfo  {journal} {Phys. Rev. D}\ }\textbf {\bibinfo {volume}
  {80}},\ \bibinfo {pages} {105005} (\bibinfo {year} {2009})}\BibitemShut
  {NoStop}%
\bibitem [{\citenamefont {{Lemm}}(1999)}]{Lemm}%
  \BibitemOpen
  \bibfield  {author} {\bibinfo {author} {\bibfnamefont {J.~C.}\ \bibnamefont
  {{Lemm}}},\ }\bibfield  {title} {\enquote {\bibinfo {title} {{Bayesian Field
  Theory: Nonparametric Approaches to Density Estimation, Regression,
  Classification, and Inverse Quantum Problems}},}\ }\href@noop {} {\bibfield
  {journal} {\bibinfo  {journal} {ArXiv Physics e-prints}\ } (\bibinfo {year}
  {1999})},\ \Eprint {http://arxiv.org/abs/physics/9912005} {physics/9912005}
  \BibitemShut {NoStop}%
\bibitem [{\citenamefont {Lax}\ and\ \citenamefont {Richtmyer}(1956)}]{Lax}%
  \BibitemOpen
  \bibfield  {author} {\bibinfo {author} {\bibfnamefont {P.~D.}\ \bibnamefont
  {Lax}}\ and\ \bibinfo {author} {\bibfnamefont {R.~D.}\ \bibnamefont
  {Richtmyer}},\ }\bibfield  {title} {\enquote {\bibinfo {title} {Survey of the
  stability of linear finite difference equations},}\ }\href {\doibase
  10.1002/cpa.3160090206} {\bibfield  {journal} {\bibinfo  {journal}
  {Communications on Pure and Applied Mathematics}\ }\textbf {\bibinfo {volume}
  {9}},\ \bibinfo {pages} {267--293} (\bibinfo {year} {1956})}\BibitemShut
  {NoStop}%
\bibitem [{\citenamefont {Corless}\ and\ \citenamefont
  {Fillion}(2013)}]{Numerics}%
  \BibitemOpen
  \bibfield  {author} {\bibinfo {author} {\bibfnamefont {R.}~\bibnamefont
  {Corless}}\ and\ \bibinfo {author} {\bibfnamefont {N.}~\bibnamefont
  {Fillion}},\ }\href@noop {} {\emph {\bibinfo {title} {A Graduate Introduction
  to Numerical Methods: From the Viewpoint of Backward Error Analysis}}},\
  SpringerLink : B{\"u}cher\ (\bibinfo  {publisher} {Springer New York},\
  \bibinfo {year} {2013})\BibitemShut {NoStop}%
\bibitem [{\citenamefont {Hirsch}(2007)}]{Godunov}%
  \BibitemOpen
  \bibfield  {author} {\bibinfo {author} {\bibfnamefont {C.}~\bibnamefont
  {Hirsch}},\ }\href@noop {} {\emph {\bibinfo {title} {Numerical Computation of
  Internal and External Flows: Fundamentals of Computational Fluid
  Dynamics}}},\ Butterworth-Heinemann\ (\bibinfo  {publisher}
  {Elsevier/Butterworth-Heinemann},\ \bibinfo {year} {2007})\BibitemShut
  {NoStop}%
\bibitem [{\citenamefont {Reed}\ and\ \citenamefont {Simon}(1975)}]{Reedsimon}%
  \BibitemOpen
  \bibfield  {author} {\bibinfo {author} {\bibfnamefont {M.}~\bibnamefont
  {Reed}}\ and\ \bibinfo {author} {\bibfnamefont {B.}~\bibnamefont {Simon}},\
  }\href@noop {} {\emph {\bibinfo {title} {Methods of Modern Mathematical
  Physics}}},\ \bibinfo {series} {Methods of Modern Mathematical Physics}\ No.\
  \bibinfo {number} {v. 2}\ (\bibinfo  {publisher} {Academic Press},\ \bibinfo
  {year} {1975})\BibitemShut {NoStop}%
\end{thebibliography}%

\end{document}